\documentclass[amsmath,amssymb,aps,pra,twocolumn]{revtex4-1}
\usepackage[T1]{fontenc}
\usepackage{amsmath}
\usepackage{amsfonts}
\usepackage{amssymb}
\usepackage{braket}
\usepackage{bbold}
\usepackage{physics}
\usepackage{tikz}
\usepackage{graphicx}
\usepackage{verbatim}
\usepackage{dcolumn}
\usepackage{bm}
\usepackage[caption=false]{subfig}

\newcommand{\sign}{\mathop{\mathrm{sign}}}

\begin{document}

\preprint{APS/123-QED}

\title{Superradiant Quantum Phase transition\\
for Landau Polaritons with Rashba and Zeeman couplings}

\author{Guillaume Manzanares}
 \author{Thierry Champel}
 \author{Denis M. Basko}
\author{Pierre Nataf}
\affiliation{
 Univ. Grenoble Alpes, CNRS, LPMMC, 38000 Grenoble, France}

\date{\today}

\begin{abstract}
We develop a theory of cavity quantum electrodynamics for a two-dimensional electron gas in the presence of Rashba spin-orbit and Zeeman couplings and perpendicular magnetic field, coupled to a spatially nonuniform 
quantum  photon field. We show that the superradiant quantum phase transition (SQPT), also known as photon condensation, can in principle occur through a pure in-plane Zeeman coupling, but it requires extremely small (unrealistic) quantum well widths or extremely fine tuning of the effective Land\'e factor which makes two Landau levels coincide. Landau level crossings can also be induced by the Rashba spin-orbit coupling and they promote the SQPT which can be obtained for certain values of the effective Land\'{e} factor and filling factors. 

\end{abstract}

\maketitle

\section{Introduction}
Cavity quantum materials \cite{schlawin2021cavity} are becoming an emergent field which bridges collective many-body phenomena in solid-state devices with strong light-matter coupling in cavity quantum electrodynamics (CQED). Among the paradigmatic models of CQED, the Dicke model \cite{Dicke_1954}  describes the interaction of $N$ identical two-level systems with a single photonic mode of the cavity.
The interaction between the atomic transition and the cavity field is measured by the vacuum Rabi frequency $\Omega_0$. In the regime with $\Omega_0$ comparable to the atomic transition frequency (the so-called  ultrastrong coupling regime~\cite{Ciuti_2005}), and for a large number of atoms coupled to the same cavity mode, a superradiant quantum phase transition (SQPT) has been predicted \cite{Emary_2003,Hepp_1973}.
It has been shown that in the thermodynamic limit, 
 the lower polariton mode exhibits a gapless critical point which separates two phases, the normal and the superradiant phases. In the superradiant phase the ground state  is characterized by a finite static average of the photon field.
A related transition has been observed \cite{esslinger_2010} in  a driven-dissipative quantum simulator of the Dicke model, but it is physically different from equilibrium superradiance \cite{Kirton2019}.

To the best of our knowledge, the SQPT has never been observed in a physical matter system coupled to the electromagnetic field although the ultrastrong coupling regime has been reached in a two-dimensional electron gas (2DEG) placed in a cavity and subject to a perpendicular static magnetic field. In this system where the matter excitations are represented by the cyclotron resonance \cite{Hagenmuller_2010,Scalari_2012}, a softening of the lowest (Landau) polaritonic excitation branch has been reported \cite{keller2020landau}. At the theoretical level, the Dicke model breaks gauge invariance and thus one needs to extend the physical model by taking into account other terms  such as the ${\bf A}^2$ term coming from the minimal coupling replacement (here ${\bf A}$ is the vector potential). However, this term is responsible for the disappearance of the SQPT for
uniform photonic field (due to gauge invariance), a result expressed by so-called ``No-go theorems'' \cite{rzazewski1975phase,polonais_1979,polonais_1981,nataf_nogo_2010,Todorov_2012,Hayn_2012,
polini_2012,Bamba_2014,rousseau_2017,andolina_nogo}. These theorems guarantee that a static spatially uniform vector potential cannot be an order parameter of the superradiant phase, since it can always be eliminated by a gauge transformation.

The order parameter distinguishing the superradiant phase must be a gauge-invariant quantity, such as the electric or magnetic field. In the former case, the SQPT is essentially driven by the Coulomb interaction, and upon a proper microscopic treatment the SQPT assumes the more common shape of a crystallization \cite{Vukics_2015} or a ferroelectric \cite{Keeling_2007,DeBernardis_2018} instability. If one looks for the SQPT driven by the transverse photonic field, the order parameter is the magnetic field, necessarily associated with a spatially non-uniform vector potential.
Several proposals in this direction have been made, including systems with magnetic-dipole interactions due to the cavity magnetic fields \cite{Knight_PRA,Andolina2020,roman2021photon} (which will play a key role in the following), or its circuit QED analog with an inductive coupling \cite{nataf_2011,nataf_2010}. More recently, a magnonic SQPT \cite{bamba2022magnonic} has been predicted where the role of the photons is played by magnons. In the same spirit as Ref. \cite{Knight_PRA}, a system of magnetic molecules coupled to a microwave cavity via the Zeeman interaction \cite{roman2021photon,jenkins2013coupling} can undergo the equilibrium superradiant phase transition.

As pointed out in Ref.~\cite{Nataf2019} and further generalized in Ref.~\citep{Andolina2020}, the SQPT identified by the magnetic field order parameter can also be viewed as a more familiar paramagnetic instability (a well-known example of which is Condon domains~\cite{ShoenbergBook}), and described in terms of a paramagnetic susceptibility which should exceed a certain critical value which depends on the specific geometry. The transition is then driven by magnetostatic interactions, which are typically rather weak, so fine tuning of parameters is required to reach the required value of the susceptibility.
In Ref.~\cite{Nataf2019}, the susceptibility of a 2DEG under a perpendicular magnetic field was found to be enhanced near Landau level crossings which occur in the presence of a sufficiently strong Rashba spin-orbit coupling. In Refs.~\cite{Guerci2020, Guerci2021, Sanchez2021}, van Hove singularities were exploited to enhance the susceptibility. In all these cases, the instability region in the parameter space was extremely narrow.

Having in mind the goal of increasing the instability region, in the present work we study theoretically the Landau polariton system in some detail with additional physical ingredients with respect to Ref. \cite{Nataf2019}. Firstly, the quantum well hosting the 2DEG is located inside the cavity at an arbitrary position, what provides the opportunity to vary the amplitude of the ${\bf A}^2$ term. Furthermore, we consider that the electronic motion in the 2DEG plane is subject to both Zeeman and Rashba spin-orbit couplings (note that throughout this paper we neglect Coulomb interaction effects). We first show that a superradiant instability may in principle occur in this system as a result of the paramagnetic nature of the Zeeman interaction only (i.~e., without Rashba spin-orbit coupling). 
This mechanism of SQPT is driven by the in-plane component of the photon magnetic field concentrated inside the quantum well. Yet, we find that the instability condition is {\it almost always} reached for unrealistically small quantum well widths.
However, it appears that even for pure Zeeman interaction, there exists specific values of the effective Land\'{e} g-factor which makes two Landau levels coincide, corresponding to magnetic spin flip transitions with no energy price.
Then, the superradiant instability is boosted and can occur for realistic quantum well width providing that the Land\'e factor is fine-tuned.

Moreover, as previously pointed out in Ref. \cite{Nataf2019}, another SQPT mechanism, associated with an out-of-plane component of the photon magnetic field, spatially modulated with a typical in-plane wave vector set by the inverse cyclotron radius, takes place in the presence of the Rashba spin-orbit coupling only. Within this mechanism, the instability is then stimulated at certain values of the applied perpendicular magnetic field by the crossings of the Landau levels corresponding also to dipole-allowed excitations with zero energy. The presence of such intrinsic soft excitations greatly enhances the effect of the coupling to the photon field. When taking into account a Zeeman
interaction on top of the spin-orbit coupled 2DEG system the Landau levels still cross (albeit for different typical values of the applied field which are
determined by the Zeeman and Rashba coupling amplitudes), so that the softening of the excitations leading to the appearance of a SQPT remains. With the
developed Rashba-Zeeman cavity QED theory, we find that the SQPT instability is promoted either for a zero or a finite value of the in-plane wave vector of the perpendicular photon field depending on  the effective Land\'{e} g-factor, the  value of the filling factor, and on the 2DEG position in
the cavity. The latter conditions determine which one of the two above different instability mechanisms  dominates. 

The paper is organized as follows. In Sec. \ref{analytique}, we introduce the model and provide the main equations allowing the determination of the polaritonic excitations in the presence of both Zeeman and Rashba couplings. 
In Sec. \ref{Superradiant Zeeman}, we analyze in a first stage the SQPT mechanism arising in the case of an in-plane Zeeman coupling. 
Then, in Sec. \ref{both} we consider the interplay of Zeeman and Rashba interaction couplings, and present a detailed study of the instability regions in the parameter space. Some technical details are provided in two Appendices.

\section{Model and analytical results}

\label{analytique}

\subsection{Model}
We consider a quantum well hosting a 2DEG with the single-electron Hamiltonian containing Rashba and Zeeman coupling terms:
\begin{eqnarray}
H 
& =& \frac{1}{2 m_{\ast}}  \left( {\bf p} +  \frac{e}{c}  {\bf A}
\right)^2  + \alpha \left({\bf p} + \frac{e}{c} {\bf A}
 \right)\times {\boldsymbol \sigma} \cdot {\bf u}_z \nonumber \\
& & {} + \frac{ \mu_B}{2}  \mathbf{B}
\cdot \hat{g} \cdot {\boldsymbol \sigma}.
\label{2DEG}
\end{eqnarray}
Here ${\bf p} = - i \hbar (\partial_x,\partial_y)$ is the 2D in-plane electron momentum, $m_{\ast}$ is the effective electronic mass, $\boldsymbol{\sigma} = (\sigma_x,\sigma_y,\sigma_z)$ is
the vector of Pauli matrices, $\alpha$ is the Rashba spin-orbit
coupling constant, and $-e < 0 $ is the electron charge.
$\mathbf{u}_z$ is the unit vector in the $z$ direction.
To account for the anisotropy of the Zeeman interaction (with $\mu_B$ the Bohr's magneton in prefactor), we have introduced a tensor  $\hat{g}$
for the effective Land\'{e} factor assuming different values for the in-plane and out-of plane components:
\begin{align}
 \hat{g}=  \begin{pmatrix}
g_{\parallel} & 0 & 0 \\
0 & g_{\parallel} & 0 \\
0 & 0 & g_{\perp}
 \end{pmatrix}.
\end{align}
The magnetic field consists of two parts, $\mathbf{ B}
=\mathbf{B}_{\mathrm{ext}}+\mathbf{B}_{\mathrm{cav}}$, where
$\mathbf{B}_{\mathrm{ext}}  = B {\bf u}_{z}  $ corresponds to an external magnetic field applied perpendicularly to the 2DEG, while
 $\mathbf{B}_{\mathrm{cav}}$  refers to the cavity electromagnetic field (a similar notation is used for the associated vector potentials).

\begin{figure}[t]
\includegraphics[scale=0.4]{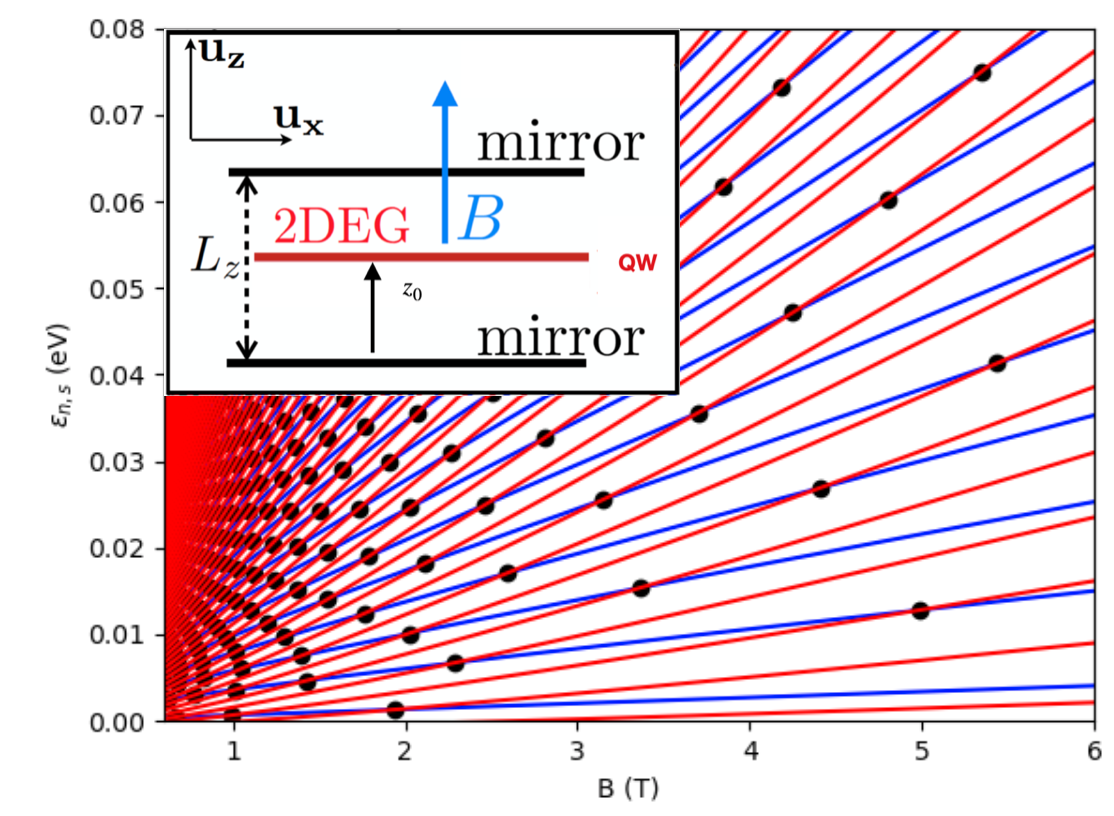}
\caption{LL energies  $\epsilon_{n,\lambda}$ [Eq. \eqref{energies propres}] versus $B$ for an effective Land\'e factor $g_{\perp} =1.26 $, an effective electron mass $m_{\ast} = 0.081$ $m_e$ and a Rashba spin-orbit coupling $\alpha =0.37$ eV.\text{\AA{}}. 
The blue (red) curves correspond to $\lambda =+1$ ($\lambda =-1$). The black dots mark the crossings given by Eq. \eqref{croisement}. The inset shows the system geometry.}
\label{Figure1}
\end{figure}

In the presence of  $\mathbf{B}_{\mathrm{ext}}$ only, the single-electron spectrum of the 2DEG is characterized by Landau levels (LL)
\begin{align}
\label{energies propres}
\epsilon_{n, \lambda } = \hslash \omega_c \left[n + \frac{\lambda}{2}  \sqrt{ \left(1- Z \right)^2 + 8n \left(\frac{m_{\ast}\alpha{l}_B}{\hbar}\right)^2 } \right],
\end{align}
where $n $ is the LL index and $ \lambda =\pm $ is the pseudo-spin index for $n \geq 1$ ($\lambda=+1$ for $n=0$). Here, $\omega_c = eB/(m_{\ast}c )$ and $ Z  = g_{\perp}  m_{\ast} /(2 m_\mathrm{e}) $ (with $m_\mathrm{e}$ the free electron mass). Each LL has a degeneracy $ L_x L_y/(2 \pi l_B^2)$ where $L_x L_y$ is the sample area and $l_B = \sqrt{\hbar c/(e B)}$ is the magnetic length. We shall assume to be at zero temperature, at a fixed electron density $n_e$, and at an external magnetic field $B$ corresponding to an integer filling factor $\nu = 2 \pi l_B^2 n_e$.

In Fig. \ref{Figure1} we have
plotted the energy levels $\epsilon_{n,\lambda}$ as a function of the magnetic field $B$ for parameters relevant
 for the material  InP (see e.g. Ref. \cite{Hermann1977}).
 As clearly seen, a characteristic feature of the spectrum is the presence of level crossings (showcased by the dots in Fig. \ref{Figure1}) between LLs $(n_1,+)$ and $(n_2,-)$ with $n_1<n_2$, which correspond to special values of the magnetic field, of the $g_{\perp}$ factor, or of  the
 Rashba spin-orbit amplitude $\alpha$ given by the conditions \cite{Champel2013}
 \begin{eqnarray}
\label{croisement}
2\left(\frac{m_{\ast}\alpha{l}_B}{\hbar}\right)^2 = n_1+n_2- \sqrt{ (1- Z)^2 + 4 n_1 n_2 }, 
\end{eqnarray}
provided that  $\vert n_2 - n_1 \vert \geq \vert 1 - Z \vert$.

As illustrated in the inset of Fig. \ref{Figure1}, we consider that the 2DEG is placed inside the optical cavity at the vertical position $z_0$. The vector potential $ \bf{A}_{\mathrm{cav}}(\bf{r})$ of the photonic field is defined by the mode expansion, determined by the cavity shape. Like in Refs. \cite{Hagenmuller_2010,Nataf2019}, we assume a perfect metallic cavity with dimensions $L_x\gg L_z\gg L_y $, filled by a material with a dielectric constant $\epsilon$, with the tangential components of the electric field, and thus of the vector potential, vanishing at the mirrors. Thus, we can take into consideration only the resonator modes with the wave vector $ {\bf q} = (q_x,0,q_z) $, where $q_x$ is continuous and $q_z =\pi n_z /L_z $ with $n_z $ a positive integer. The corresponding mode frequencies are $\omega_{q_x,n_z} = (c/\epsilon) \sqrt{q_x^2 +q_z^2}$.
In this case, the cavity vector potential reads \cite{kakazu_kim}
\begin{eqnarray}
\label{Eq1}
{\bf A}_{\mathrm{cav}}({\bf r})
&=&  {\bf u}_y \sum_{q_x,n_z} \sqrt{\frac{4 \pi \hbar c^2}{L_xL_yL_z \epsilon \omega_{q_x,n_z}}}  \sin(\frac{n_z \pi z}{L_z}) \nonumber \\
&& \times
 \left(a_{q_x,n_z} e^{i q_x x} + a_{q_x,n_z}^{\dagger} e^{-i q_x x} \right),
\end{eqnarray}
where $a_{q_x,n_z}^{\dagger}$ ($a_{q_x,n_z}$) is the photon creation (annihilation)
operator and ${\bf u}_y$ is the unit vector in the $y$ direction. 

In the discussion above we implicitly assumed the 2DEG to be infinitely thin, so the vector potential $\mathbf{A}$ and the magnetic field $\mathbf{B}$ entering Eq.~(\ref{2DEG}) are taken at $z=z_0$.
In fact, any confining potential in the $z$ direction gives rise to multiple electronic subbands; here we assume that only the lowest one is occupied, so all electronic wave functions are proportional to $\zeta(z)$, the wave function of this lowest subband. Then, what enters Eq.~(\ref{2DEG}), are in fact the convolutions $\int\mathbf{A}(\mathbf{r})\,\zeta^2(z)\,dz$ and  $\int\mathbf{B}(\mathbf{r})\,\zeta^2(z)\,dz$. The assumption of an infinitely thin quantum well corresponds to $\zeta^2(z)\to\delta(z-z_0)$; we will see however that the results also contain the integral $\int\zeta^4(z)\,dz\equiv1/W$, which will be our definition of the quantum well width~$W$. Our assumption is that this is the smallest length scale in the problem (in particular, $W\ll1/q_x,L_z$).

\subsection{Polariton modes}

The SQPT is signaled by the vanishing of the lowest polariton frequency. The polariton modes, which refer to the excitations of the coupled
2DEG-cavity system, can be found by several different methods, see  Ref. \cite{Nataf2019} for instance.
  Typically, they correspond to  the non-zero solutions of the Amp\`{e}re-Maxwell differential law
\begin{align}
\label{eq}
\left[ \epsilon \frac{\omega^2}{c^2} + \nabla^2\right] A_i(\mathbf{r}) =  \frac{4  \pi}{c^2}  \int \mathrm{d} \mathbf{r}'\,\mathcal{Q}_{ij }(\mathbf{r},\mathbf{r}', \omega) A_j(\mathbf{r}')
\end{align}
with ${\bm \nabla} \cdot {\bf A}=0$.
The source current term contains the 2DEG response function $\mathcal{Q}_{ij }$, which determines the response of the current density $ \delta \bf{j}$ to a change in the vector potential $ \bf{A} = \bf{A}_{\mathrm{ext}} + \delta\bf{A}$ in the linear order
\begin{align}
\label{deltaj}
\delta j_{k} ({\bf r}, \omega) = -  \int  \mathrm{d^3}{\bf r'}~  \sum_{l} \mathcal{Q}_{kl}({\bf r},{\bf r'}, \omega) \frac{\delta A_{l}({\bf r'}, \omega) }{c}. \end{align}
 In the following, due to the geometry of the problem, we will focus on the response function in the $y$ direction, $\emph{i.e}$, $\mathcal{Q}_{yy}({\bf r},{\bf r'}, \omega) $.
Let us call  $ Q({ \bf q}_{\|},z,z', \omega) $ the Fourier transform of  $\mathcal{Q}_{yy} $ at ${\bf q}_{\|} = q_x {\bf u}_x $:
\begin{align}
Q(q_x,z,z', \omega) &= \int \mathrm{d} {\bf r}_{\|} ~e^{-i q_x (x-x')} \mathcal{Q}_{yy}({\bf r},{\bf r'}, \omega),
\label{TF}
\end{align}
where the  index ${\|}$  refers to the components along the $x$ and $y$ axis.  As shown in Appendix \ref{AppA}, for an infinitely thin quantum well we can decompose $Q(q_x,z,z', \omega)$
into four terms:
\begin{eqnarray}
\label{susceptibilite}
Q(q_x,z,z', \omega)&=& - Q_{0}(q_x,\omega)\, \delta(z-z_0 )\, \delta(z'-z_0 ) \nonumber\\
&&-   Q_{1}(q_x,\omega)\, \delta(z-z_0 )\, \delta'(z'-z_0 ) \nonumber \\
&& - Q_{1}(q_x,\omega)\, \delta'(z-z_0 )\, \delta(z'-z_0 )  \nonumber \\
&& -  Q_{2}(q_x,\omega) \, \delta'(z-z_0  ) \,\delta'(z'-z_0  )
\end{eqnarray}
with
\begin{eqnarray}
\label{Q0}
Q_{0}(q_x,\omega) & =&   \frac{1 }{  \pi l_B^2} \sum_{ \ell \leq \nu < \ell'} \frac{(\epsilon_{\ell} - \epsilon_{\ell'})
 \left(\mathcal{A}_{q_x}^{\ell' \ell}\right)^2 }{(\hbar \omega)^2 -(\epsilon_{\ell'} - \epsilon_{\ell})^2 } - \frac{ n_{e} e^2}{m_{\ast}},  \hspace*{0.5cm}  \\
Q_{1}(q_x,\omega) &= &\frac{1 }{  \pi l_B^2} \sum_{ \ell \leq \nu < \ell'} \frac{(\epsilon_{\ell} - \epsilon_{\ell'}) \mathcal{B}_{q_x}^{\ell' \ell} \mathcal{A}_{q_x}^{\ell' \ell}}{(\hbar \omega)^2 -(\epsilon_{\ell'} - \epsilon_{\ell})^2 } , \label{Q1} \\
Q_{2}(q_x,\omega)  & = &\frac{1 }{  \pi l_B^2} \sum_{ \ell \leq \nu < \ell'} \frac{(\epsilon_{\ell} - \epsilon_{\ell'})
\left(\mathcal{B}_{q_x}^{\ell' \ell} \right)^2 }{(\hbar \omega)^2 -(\epsilon_{\ell'} - \epsilon_{\ell})^2 } .      \label{Q2}
\end{eqnarray}
Here, the LL indices $(n,\lambda)=\ell$ are combined into a single label and ordered according to the LL energies $\epsilon_{\ell}$ given by Eq. \eqref{energies propres}, such that energy levels with $\ell \leq \nu$ are filled and those with  $\ell' > \nu$  are empty.
The quantities $\mathcal{A}_{q_x}^{\ell' \ell} $   and  $ \mathcal{B}_{q_x}^{\ell' \ell} $  appearing  in Eqs. \eqref{Q0}--\eqref{Q2} represent dipole matrix elements given by the expressions
\begin{eqnarray}
\mathcal{A}_{q_x}^{\ell' \ell} &=&   \frac{e l_B \omega_c }{\sqrt{2}}
\left[
  \left(\sqrt{n}\,\Theta_{n'}^{n-1}  - \sqrt{n+1}\, \Theta_{n'}^{n+1}\right)\cos\theta_{\ell'} \cos\theta_{\ell}\right.
   \nonumber      
\\ && \qquad{} + \left.
 \left( \sqrt{n-1}\,\Theta_{n'-1}^{n-2}  -  \sqrt{n}\,\Theta_{n'-1}^{n} \right)\sin\theta_{\ell'} \sin\theta_{\ell}\right]
 \nonumber \\
&&{} -  e\alpha \left(\Theta_{n'-1}^{n}\sin\theta_{\ell'} \cos\theta_{\ell} -  \Theta_{n'}^{n-1}\cos\theta_{\ell'} \sin\theta_{\ell}\right) \nonumber \\
&&
{}+ \frac{ e \hbar q_x}{m_{\ast}}\,\frac{Z}{2}  \left( \Theta_{n'}^{n}\cos\theta_{\ell}  \cos\theta_{\ell'} -   \Theta_{n'-1}^{n-1} \sin\theta_{\ell}  \sin\theta_{\ell'}\right), \nonumber \\
&&  \label{defA}  \\
\mathcal{B}_{q_x}^{\ell' \ell}  &=& \frac{g_{\parallel}  \mu_B c }{2} \left(   \Theta_{n'-1}^{n}\sin\theta_{\ell'}\cos\theta_{\ell} -  \Theta_{n'}^{n-1}\cos\theta_{\ell'} \sin\theta_{\ell} \right), \nonumber \\
&& \label{defB}
\end{eqnarray}
with  $\tan\theta_{l} = \lambda \sqrt{u_n^2 +1}  - u_n $ and $u_n = \hbar(1- Z)/(m_{\ast}\alpha{l}_B \sqrt{8n})$. The overlap functions $\Theta_{n_2}^{n_1}$ containing the $q_x$ dependence are given by
\begin{eqnarray}
 \Theta_{n_2}^{n_1}  = S_{n_2}^{n_1} \sqrt{\frac{m!}{M!}} \left|\frac{l_Bq_x}{\sqrt{2}} \right|^{M-m} \!\! L_{m}^{(M-m)} \left(\frac{l_B^2q_x^2}{2} \right)  e^{- \frac{l_B^2 q_x^2}{4} }   , \hspace*{0.5cm} \label{Theta}
\end{eqnarray}
with $L_{m}^{(M-m)}(x)$ the generalized Laguerre polynomial of degree $m=\mathrm{min}(n_1,n_2)$, $M=\mathrm{max}(n_1,n_2)$ and $S_{n_2}^{n_1} = \sign\left[ q_x (n_2-n_1) \right] ^{n_2-n_1}$. Note that at $q_x =0$ we have $ \Theta_{n_2}^{n_1} = \delta_{n_1,n_2}$, thus implying that the reduced coupling constants $\mathcal{A}_{q_x =0}^{\ell' \ell}$ and $\mathcal{B}_{q_x=0}^{\ell' \ell} $ are non-zero only between consecutive LLs, $n' = n \pm 1$, with no restriction on $\lambda$. At finite $q_x$, this selection rule is relaxed.

The different terms appearing in Eq. \eqref{susceptibilite} have different physical origins, which are rather explicit when looking at the coupling constants entering into the expressions of the dipole matrix elements. 
The contribution $Q_{0}$ depending only on $\mathcal{A}_{q_x}^{\ell' \ell} $ results from the coupling of the electronic charge to the perpendicular magnetic field and from the Zeeman coupling along the $z$-axis. Therefore, it typically characterizes the overall effect of an out-of-plane magnetic field.
In contrast, the quantity $Q_{2}$ only depends via $ \mathcal{B}_{q_x}^{\ell' \ell}$ on the component $g_{\parallel}$ of the Land\'{e} factor tensor, and can thus be directly related to the
effect of an in-plane magnetic field. Indeed, the $\delta$~function derivative $\delta'(z'-z_0)$ (cf Eq. \eqref{susceptibilite}) indicates that the system responds to $-\partial{A}_y/\partial{z}=B_x(z_0)$; the fact that the responding current $j_y(z)\propto\delta'(z-z_0)$ corresponds to the in-plane magnetization being $M_x(z)\propto\delta(z-z_0)$ (indeed, the current $\mathbf{j}=c\boldsymbol{\nabla}\times\mathbf{M}$, with $\mathbf{M}$ the magnetization); so $Q_2$ is nothing but the in-plane spin susceptibility.
Finally, the contribution $Q_{1}$ appears to be a mixture of the dipole matrix elements $\mathcal{A}_{q_x}^{\ell' \ell} $  and  $ \mathcal{B}_{q_x}^{\ell' \ell} $, and can be consequently seen as the result of the simultaneous presence of the in-plane and perpendicular components of the magnetic field.

In Eq. \eqref{susceptibilite}, the Dirac $\delta$ function and its derivative express the discontinuity of the vector potential component $A_y({\bf r})=A_{y}(z) \, e^{iq_x x}$
and of its derivative with respect to $z$ at the 2DEG position $z_0$. In fact, these discontinuities stem from the hypothesis of
an infinitely thin quantum well, which is crucial in order to be able to derive an analytical solution to  Eq. \eqref{eq}.
More precisely, we assume that the quantum well width is such that $\kappa W \ll 1$ where $ \kappa = \sqrt{q_x^2 - \epsilon \omega^2/c^2}$. Considering the boundary conditions $A_y(z=0)=A_y(z=L_z)=0$  imposed by the cavity geometry, we can solve Eq. \eqref{eq} for an arbitrary 2DEG position $z_0$,
see Appendix A. In the
following, we shall showcase and compare two different typical situations: (a) the 2DEG is placed in the middle of the cavity; (b) the 2DEG is placed close to a cavity mirror.  For the case (a) with $z_0= L_z/2 $, we find that the polariton frequencies are solutions of the equation:
\begin{eqnarray}
\frac{c^2}{4 \pi}  &= &
\frac{2 \pi}{c^2}   \left[ Q_{1}(q_x,\omega)^2 -  Q_{0}(q_x,\omega)  Q_{2}(q_x,\omega)  \right] \frac{\tanh(\kappa L_z/2)}{\kappa W}
    \nonumber \\
&&
+Q_{0}(q_x,\omega) \frac{\tanh(\kappa L_z/2)}{2 \kappa}  + Q_{2}(q_x,\omega)   \frac{1}{W} .      \label{Lz/2}
\end{eqnarray}
 In the case (b) with $z_0=0$, we get the different equation for the polaritonic modes:
\begin{eqnarray} \frac{ c^2}{4  \pi}   &=&
Q_{2}(q_x,\omega)  \frac{1}{W}   - \frac{\pi}{c^2}  Q_{1}(q_x,\omega)  Q_{2}(q_x,\omega) \kappa \tanh(\kappa L_z)
  \nonumber \\
&&
- \frac{\pi}{c^2}  \left[ Q_{1}(q_x,\omega)^2 - Q_{0}(q_x,\omega)  Q_{2}(q_x,\omega) \right]  -Q_{1}(q_x,\omega)
. \nonumber \\
&&
\label{z_0=0}
\end{eqnarray}
These two equations represent the main analytical result of this work. Note that Eq. \eqref{Lz/2} naturally reproduces Eq. (5) of Ref.  \cite{Nataf2019} in the absence of Zeeman coupling, {\em i.e.}, for $Q_1=Q_2=0$. In contrast, Eq. \eqref{z_0=0}, which essentially encapsulates the Zeeman interaction effect, has not been obtained previously. The subsequent sections of the paper are devoted to the physical analysis of the derived equations.


\section{Superradiant instability with Zeeman coupling}

\label{Superradiant Zeeman}
We now aim at studying the conditions for a possible softening of the polariton modes by analyzing the solutions of Eqs. \eqref{Lz/2}--\eqref{z_0=0} for $\omega=0$. The existence of such solutions signals the onset of a SQPT. The positioning of the 2DEG in the cavity has a priori an important influence. Indeed, close to the mirror ($z_0=0$ for instance) the vector potential vanishes (but not the magnetic field itself) as the result of the boundary. This allows
one to eliminate the ${\bf A}^2$ contribution in the Hamiltonian \eqref{2DEG}, which is known to have a harmful effect on the SQPT for a uniform photonic field according to No-go theorems. Let us investigate the simple situation $q_x=0$ for which the in-plane modulation $e^{iq_x x}$ of the cavity field is absent (nevertheless the magnetic field component $B_x=-\partial_z A_y$ remains nonuniform with respect to the vertical position). In this case, gauge invariance imposes the constraints $Q_0(0,0)=Q_1(0,0)=0$
(that we have checked numerically), 
so that Eqs.  \eqref{Lz/2} and  \eqref{z_0=0} boil down to an equivalent simpler equation. An instability develops as soon as
\begin{eqnarray}
\frac{Q_2(0,0)}{W} \geq \frac{c^2}{4 \pi}.
\label{cond}
\end{eqnarray}
This instability corresponds to a spontaneous generation of an in-plane magnetization $M_x\propto\delta(z-z_0)$, equivalent to a spontaneous creation of two parallel layers of opposite surface currents, $j_y/c=\partial{M}_x/\partial{z}\propto\delta'(z-z_0)$. A similar instability was found in Refs.~\cite{Guerci2021, Sanchez2021}, where two physical layers of graphene were studied. Here such two current layers are effectively produced in a single transverse subband in a quantum well by the spin in-plane magnetization.

Inequality~(\ref{cond}) is very demanding in practice because $c^2$ is much larger than the square of any velocity scale typically occurring in a solid. Obviously, the  condition \eqref{cond} calls for small quantum well widths.
In the absence of Rashba spin-orbit coupling ($\alpha=0$), the quantity $Q_2(0,0)$ can be calculated analytically. 
As a result, we can estimate the maximal (critical) value of the quantum well width yielding the SQPT instability to be (for $|Z|<1$):
\begin{align}
\label{LQW}
W_c = \frac{g_{\parallel}^2}{\vert g_{\perp}\vert}\frac{e^2}{2 m_e c^2}.
\end{align}
This equation can also be derived from a rather elementary consideration of the energy gained by a spin tilt.
Let us assume odd~$\nu$. Then all filled Landau levels are filled for both spin projections except one, which is fully spin polarized. This results in the 2DEG magnetization $\mathbf{M}(\mathbf{r})=(\sign{B}_\mathrm{ext})\mathbf{u}_z\delta(z-z_0)\,|g_\perp|\mu_B/(4\pi{l}_B^2)$. Let us now check the stability of this configuration with respect to a tilt of all spins by an infinitesimal angle~$\vartheta$. Then, to the second order in $\vartheta$, the magnetization becomes $\mathbf{M}(\mathbf{r})=(\sign{B}_\mathrm{ext})[|g_\perp|(1-\vartheta^2/2)\mathbf{u}_z+g_\|\vartheta\mathbf{u}_x]\delta(z-z_0)\mu_B/(4\pi{l}_B^2)$. The in-plane magnetic field produced by the in-plane magnetization is found from the continuity of $H_x=B_x-4\pi{M}_x$, which vanishes away from the 2DEG. Thus, $B_x(\mathbf{r})=4\pi{M}_x(\mathbf{r})=4\pi(\sign{B}_\mathrm{ext})g_\|\vartheta\mathbf{u}_x\delta(z-z_0)\mu_B/(4\pi{l}_B^2)$. The energy (per unit area)  of such configuration is
\begin{align}
&-B_\mathrm{ext}\int{M}_z(z)\,dz-\frac12\int{B}_x(z)\,{M}_x(z)\,dz={}\nonumber\\
&{}=-\frac{|g_\perp\mu_BB_\mathrm{ext}|}{4\pi{l}_B^2}
+\frac{\vartheta^2}2\left[\frac{|g_\perp\mu_BB_\mathrm{ext}|}{4\pi{l}_B^2}-\frac{4\pi}{W}\left(\frac{g_\|\mu_B}{4\pi{l}_B^2}\right)^2\right].
\end{align}
Using $\mu_B=e\hbar/(2m_\mathrm{e}c)$, we find that the tilt becomes energetically favorable if $W<W_c$ with $W_c$ given by Eq.~(\ref{cond}).
Incidentally, we also understand from this simple derivation the
presence of some products of delta functions in the susceptibility $Q$ (cf Eq. \eqref{susceptibilite}).\\
For an isotropic Zeeman interaction with $|g_{\parallel}|=|g_{\perp}|=1$, Eq. \eqref{LQW} leads to  $W_c \sim 1$ fm. Such a critical value differs by several order of magnitudes from the characteristic quantum well widths which are in the nanoscale.
 We thus conclude that the SQPT can in principle occur for $q_x=0$ via the in-plane Zeeman interaction coupling, but it is not experimentally achievable  in Landau polariton systems.



\begin{figure*}[t]
\subfloat[$z_0=0$ and $g_{\perp}=g_{\parallel}=g$ \label{Fig2a} ]{
\includegraphics[width=.32\linewidth]{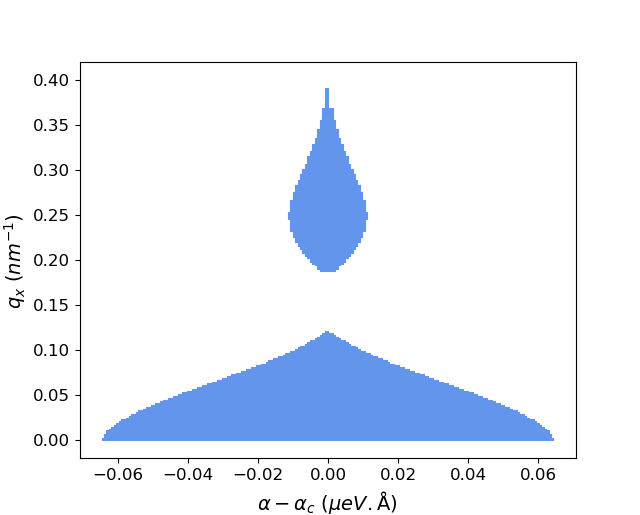} }
\hfill
\subfloat[ $z_0=L_z/2$ and $g_{\perp}=g_{\parallel}=g$  \label{Fig2b}]{
\includegraphics[width=.32\linewidth]{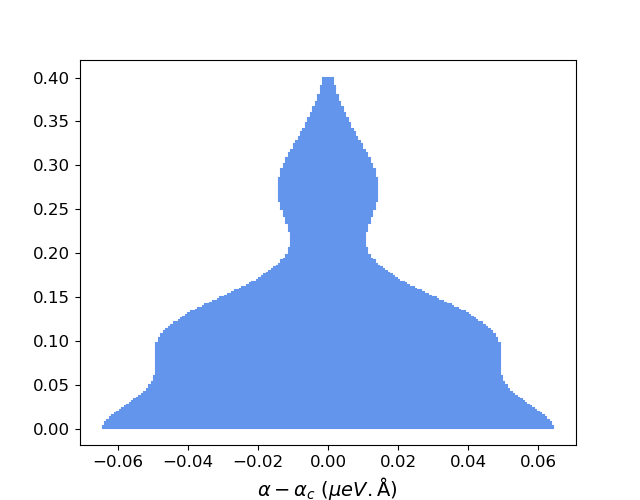}}
\hfill
\subfloat[ $z_0=L_z/2$ and $g_{\perp}=g$, $g_{\parallel}=0$
\label{Fig2c}
]{
\includegraphics[width=.32\linewidth]{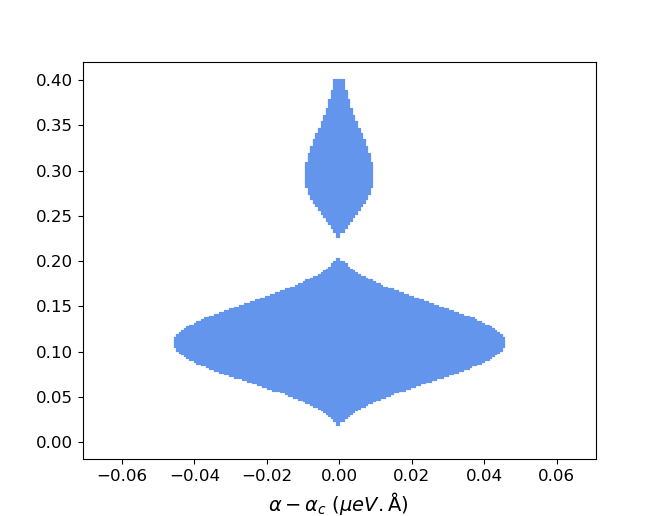}}
 \caption{Instability regions (in blue) in the parameter plane $(\alpha,q_x)$ for different 2DEG positions and 
configurations of
the $\hat{g}$ tensor.  Each region is centered around the same  $\alpha_c=0.074747587$ eV.\text{\AA{}}, which corresponds to a LL crossing selected by the choice of $\nu=3$ and $B=5$ T. Here, we fix the quantum well width $W=1$ nm, and have taken  $ g=1.26$ and
 $m_{\ast} = 0.081 \, m_e $. }
\label{Figure2}
\end{figure*}

However, it is possible to boost the amplitude of $Q_2$ by several order of magnitudes by fine tuning the effective perpendicular Land\'e factor $g_{\perp}$.
Indeed, we note with expression \eqref{Q2} that when $\omega=0$, the term $Q_2$ might diverge when there are two LLs $(n_1,+)$ and $(n_2,-)$ such that    $\epsilon_{(n_2,-)} -\epsilon_{(n_1,+)}$ vanishes with $\mathcal{B}_{q_x}^{(n_1,+)(n_2,-)} \neq 0$.
This kind of level crossing scenario is reminiscent of the mechanism producing the superradiant instability with Rashba spin-orbit coupling $\alpha \neq 0$ as studied in \cite{Nataf2019}. Surprisingly, it can also occur without Rashba coupling ($\alpha = 0$),
when $g_{\perp}$ is such that $\vert 1-Z \vert = \vert  n_1-n_2\vert $, according to Eq. \eqref{croisement}. There, the LLs $(n_1,+)$ and $(n_2,-)$ are completely superposed for all values of the magnetic field $B$.
Then, for $g_{\perp}$ close to the specific values  $g_{\perp}^c=Z^c \left (\frac{2 m_e}{m_{\ast}} \right )$, where the integer $Z^c=1 \pm \vert n_2-n_1\vert$, 
the superradiant instability can take place for a quantum well width smaller than

\begin{align}
\label{delta_Z_LQW}
W_c \sim  \frac{g_{\parallel}^2}{\vert g_{\perp}- g_{\perp}^c\vert } \frac{e^2}{2 m_e c^2}.
\end{align}
The instability is also produced here by the coupling to the in-plane component $B_x$ of the cavity magnetic field.
As detailed in Appendix \ref{AppB}, this occurs for finite wave-vector $q_x$, under conditions for which $Q_0$ and $Q_1$ play almost no role in Eqs.  \eqref{Lz/2} and  \eqref{z_0=0}.
Consequently, a fine tuning of $g_{\perp}$ can lead to an arbitrarily large upper bound of the quantum well width $W_c$.

For $g_{\perp}$ far from $g_{\perp}^c$, the divergence of $Q_2$ can also come from a LL crossing induced by the Rashba spin-orbit coupling in the 2DEG system. We still need to have finite dipole matrix elements $\mathcal{B}_{q_x}^{\ell' \ell} $ under the conditions of LL crossing $\epsilon_{(n_2,-)}=\epsilon_{(n_1,+)}$, as shown in Eq. \eqref{Q2}  for $\omega=0$. At $q_x=0$, these matrix elements are non-zero only when considering consecutive LLs $n_2=n_1 \pm 1$. According to Eq. \eqref{croisement} and the associated condition $|n_2-n_1|\geq \vert 1-Z \vert $, level crossings between consecutive levels are only possible when $g_{\perp} \geq0$. Therefore, the $Q_{2}$ boost scenario promoting the SQPT instability via the coupling to an in-plane magnetic field  $B_x$ of the photons is conceivable for reasonable $W$ at $q_x=0$ only for positive $g_{\perp}$ factors and thanks to the Rasbha spin-orbit coupling. 
 Note that an in-plane modulation $e^{i q_x x}$ of the vector potential $A_y$ producing an out-of-plane field component $B_z=\partial_x A_y$ provides another access towards the superradiant instability \cite{Nataf2019} taking place already in the absence of Zeeman coupling. At finite $q_x$, both $B_z$-driven and  $B_x$-driven instability mechanisms are in principle possible independently of the sign of $g_{\perp}$. The widening of the parameter space leads then to novel opportunities for the occurrence of the SQPT, which are studied in detail in the next section.

\section{Superradiant instability with Zeeman and Rashba couplings}

\label{both}
\subsection{Instability regions  in the parameter plane $(\alpha,q_x)$  }

From now on, we consider the general situation with $q_x \neq 0$ and the interplay of nonzero Zeeman and Rashba spin-orbit couplings. Let us first set fixed values for the quantum well width, the effective Land\'e factors, the external magnetic field and the filling factor ($W=1$ nm, $B=5$ T and $\nu=n_1+n_2=3$). In this situation, the superadiant instability arises close to level crossings, {\em i.e.}, for values of the spin-orbit coupling constant $\alpha$ close to those given by Eq. \eqref{croisement}. By numerically solving Eq. \eqref{Lz/2} or Eq. \eqref{z_0=0} for $\omega=0$, we get the boundaries of the instability regions in the $(\alpha,q_x)$ parameter plane displayed by color shading in Fig. \ref{Figure2} for different 2DEG positions and configurations of the $\hat{g}$ tensor (throughout, we consider the limit $L_z \to \infty$).
These instability regions determine the values of $\alpha$ for which the system is in the superradiant state: for a given~$\alpha$, if there is at least one value of~$q_x$ which falls into a shaded region, the system is unstable. From the shape of the shaded regions, one can also read the value of~$q_x$ at which the instability develops: it is the one that goes unstable the first, as $\alpha$ approaches the shaded regions from outside.

For a 2DEG close to one of the cavity mirrors ($z_0=0$) and an isotropic Zeeman interaction $g_\|=g_{\perp}=g=1.26$, we observe with Fig. \ref{Fig2a} that the instability first occurs
at $q_x=0$, which constitutes a distinguishing feature of the SQPT mechanism driven by the in-plane
Zeeman interaction discussed in the previous section. Clearly, for the chosen values of the  $\hat{g}$ tensor, the spatial modulation of the field is detrimental to this mechanism as illustrated by the triangular shape. As shown in Appendix \ref{AppB}, this can essentially be related to a reduction of the amplitude of the $Q_2(q_x,0)$ term
when $q_x$ increases.  If the 2DEG is instead located in the middle of the cavity ($z_0=L_z/2$), the width $\Delta \alpha$ of the instability region then exhibits a more complex (non-monotonic) dependence on $q_x$, see Fig. \ref{Fig2b}. It turns out that at $q_x \neq 0$ the contribution $Q_{0}(q_x,0)$ in Eq. \eqref{Lz/2} starts to also play a role: Both $Q_0$ and $Q_2$ terms then work together to promote the instability (note that the other contributions in the equation
are negligible close to a level crossing, see Appendix \ref{AppB}).

To better understand the complicated shape shown in Fig. \ref{Fig2b}, it is instructive to turn off the $Q_2$ contribution by setting $g_{\parallel}=0$ and keeping the other parameters unchanged. This leads to Fig. \ref{Fig2c}, which displays instability regions as bubbles. These characteristic shapes are reminiscent of those found in Ref. \cite{Nataf2019}, where the instability develops around a typical finite $q_x$  given by the inverse cyclotron radius $(\sqrt{\nu} l_B)^{-1}$. The consideration of
 $g_{\perp} \neq 0$ induces quantitative modifications for the instability regions but does not fundamentally change the instability mechanism  that was found  in Ref. \cite{Nataf2019}, unlike to the effect of the other component $g_{\parallel}$ of
 the $\hat{g}$ tensor.
We deduce that the diagram shape seen in Fig. \ref{Fig2b} can be interpreted as the superposition of Figs. \ref{Fig2a} and \ref{Fig2c}, {\em i.e.},
is the result of the coexistence of two different SQPT mechanisms in some parameter ranges.

\subsection{Critical quantum well width}

\begin{figure}[t]
\centering
\includegraphics[width=.95\linewidth]{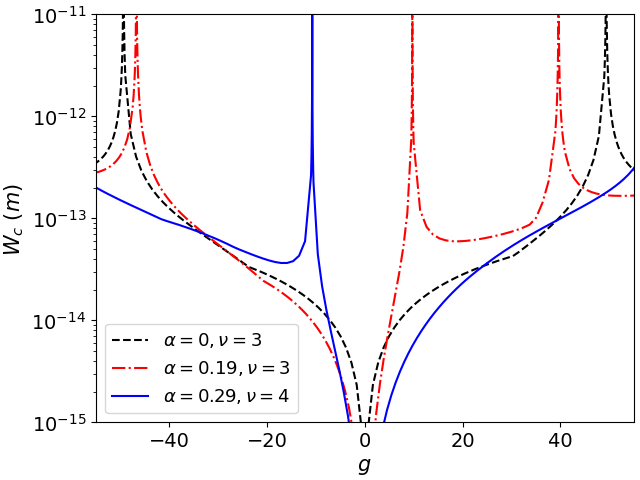}
\caption{$W_{c}$ versus $g$. In black (dashed line): $\alpha=0$ and $\nu=3$. In red (dashed-dot line):  $\alpha=0.19$ eV.\text{\AA{}} and $\nu=3$. In black (solid line): $\alpha=0.29$ eV.\text{\AA{}} and $\nu=4$. The peaks are divergences due to LL crossings or LL coincidence. See text for details.}
\label{Fig3}
\end{figure}

So far, we have worked at a fixed $W$. We now aim at revisiting the previously established
instability criterions \eqref{LQW} and \eqref{delta_Z_LQW}  on the critical quantum well width $W_c$ by taking into account  the additional effects of a spin-orbit coupling.  The 2DEG is held at the fixed position $z_0=0$ and the Zeeman coupling is taken to be isotropic $g_{\perp}=g_{\parallel}=g$. In Fig. \ref{Fig3}, we show the dependence of  $W_c$ determined numerically from Eq. \eqref{z_0=0} as a function of $g$ for different values of $\alpha$ and of the filling
factor $\nu$ (we again consider $L_z \to \infty$ and $B=5$ T). As a reference case,  the (black) dashed-line corresponds to the result obtained for $\alpha=0$
and $\nu=3$.  It perfectly
corresponds to the analytical results of Sec. \ref{Superradiant Zeeman}, i.e. Eq. \eqref{LQW} for $g$ around $0$, where $W_c$ vanishes linearly with $g$  and Eq. \eqref{delta_Z_LQW}  for $g$ close to $g_{\perp}^c= \pm 4 m_e/m_{\ast}$ (i.e $Z^c=\pm 2$ in the displayed range of $g$), where $W_c$ diverges like $1/\vert g-g_{\perp}^c \vert $.  It also confirms that the most favorable situation
for the instability is associated with a typical value $q_x=0$ (resp. $q_x$ finite) for $g$ close to $0^+$ (resp. $g_{\perp}^c = \pm 4 m_e/m_{\ast}$).

As seen with the (red) dashed-dot line in Fig. \ref{Fig3}, a nonzero $\alpha$ tends to promote the instability at $\nu=3$ in a large range of 
positive $g$, where the $W_c$ value gets enhanced in comparison to the $\alpha=0$ case. The highest $W_c$ is associated here to an absence of modulation ($q_x=0$), thus indicating that the $B_x$-instability mechanism is  the dominant one for the considered parameters. For values of $g$ approaching the LL crossing condition \eqref{croisement}, the divergences of $W_c$ prove that the instability may even
develop for $W$ in the nanoscale (but only for a very fine-tuned value of $g$ as manifested by the extremely 
sharp peaks), fully consistent with the findings of Fig. \ref{Fig2a} where $W$ was pinned to the value of 1~nm.
The different peaks correspond to different sets of LLs $(n_1,n_2)$ satisfying the LL crossing condition \eqref{croisement} with $\nu=n_2+n_1$.
For a different filling factor, for instance $\nu=4$, which corresponds to the result shown with the (blue) solid line of Fig. \ref{Fig3}, the main features are very similar to that for $\nu=3$. However, the largest $W_c$ values are systematically obtained now for $q_x \neq 0$, thus indicating that the dominant instability mechanism at play in this case is the one involving the $e^{iq_x x}$ modulation of the cavity magnetic field (situation close to that 
seen in Fig. \ref{Fig2c}). This difference with respect to the $\nu=3$ case originates from the different possible selection rules associated to the relevant transitions (see Appendix \ref{AppB} for some technical details).

 \subsection{Evolution of $\Delta \alpha$ versus $g$ and $W$}

\begin{figure}[t]
\centering
 \includegraphics[width=1\linewidth]{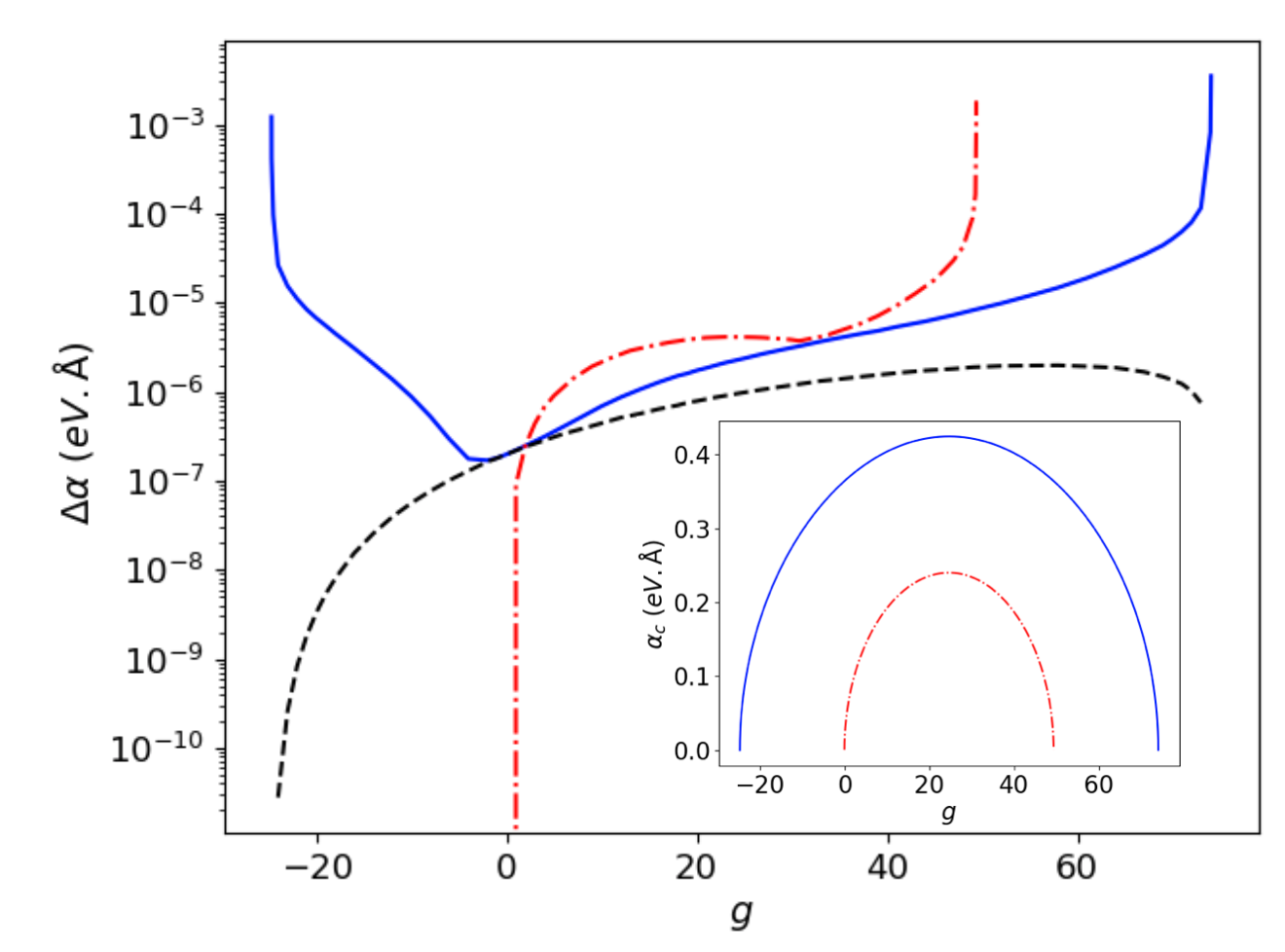}
\caption{$\Delta\alpha$ versus $g$. Here $W =1 $ nm, $m_{\ast}= 0.081 m_e$ and $z_0 =L_z/2$. In black (dashed line),   $\nu=4$ and $g_{\parallel} =0$ so $g_{\perp}=g$. For the two other curves (in dashed-dot red $\nu=3$, in solid blue $\nu=4$),
$g_{\perp}=g_{\parallel}=g$:  There is no divergence but rather  peaks which saturate around $10^{-3}$ for values of $g$ corresponding to $\alpha_c=0$. Inset: Dependence of $\alpha_c$ on $g$ for $\nu=3$ (in red) and $\nu=4$ (in blue), as given by Eq. \eqref{croisement}. }
\label{Fig4}
\end{figure}

We can notice with Figs. \ref{Figure2} and \ref{Fig3} that the superradiant regions centered around a level crossing are very 
narrow. We wish to study now the influence of the Zeeman
interaction on the typical widths $\Delta \alpha$ of these regions. The objective is to determine the optimal material conditions for revealing the SQPT by varying the Rashba coupling constant $\alpha$, which can usually be adjusted in-situ by applying a perpendicular electric field.  Let us first fix the quantum well width $W=1$ nm, and work at $B=5$ T. 
In case 
of $g_{\parallel} =0 $ and $z_0=L_z/2$, the quantity $\Delta \alpha$ (centered around the value $\alpha_c$, which varies as a function of $g$ according to Eq. \eqref{croisement})
is plotted as a function of $g=g_{\perp}$ in black (dashed line) for $\nu=4$ in Fig. \ref{Fig4}. Such quantity depends only on the term $Q_0(q_x,0)$ governing the $B_z$-instability mechanism, since we have $Q_1=Q_2=0$ for this parameter choice. The gain
in $\Delta \alpha$ found for $g \neq 0$  in comparison to the $g=0$ case proved to be quite modest. This happens because the perpendicular Zeeman coupling leads to a small modification in the electronic spectrum and on the dipole matrix elements $\mathcal{A}_{q_x}^{\ell' \ell} $.
Note that positioning the 2DEG close to the mirror ($z_0=0$) does not 
help here, since a SQPT instability is then impossible (this corresponds to $Q_1=Q_2=0$ in Eq. \eqref{z_0=0}).  Furthermore, 
this result for $\Delta \alpha$
will not be affected by a modification of the quantum well width, as long as $g_{\parallel}= 0$.

The situation with $g_{\parallel} \neq 0$ turns out to be more interesting. A configuration with an isotropic Zeeman 
coupling $g_{\parallel} =g_{\perp} =g$ now yields a more important dependence for $\Delta \alpha$ on $g$, as depicted by the red (dashed-dot) and blue (solid) curves of Fig. \ref{Fig4} corresponding to two different filling factors ($\nu=3$ in red and $\nu=4$ in blue). Clearly, the presence of 
the contribution $Q_2$ encapsulating the $B_x$-instability mechanism brings about 
a quantitative change, since a gain for $\Delta \alpha$ of several orders of magnitude 
is possible. It occurs when $g$ gets close to $g_{\perp}^c=Z^c \left (\frac{2 m_e}{m_{\ast}} \right )$, where the integer $Z^c$ depends on $\nu=n_2+n_1$ and corresponds to the perfect coincidence of LLs energies $\epsilon_{(n_2,-)}=\epsilon_{(n_1,+)}$ without Rashba coupling studied in the section \ref{Superradiant Zeeman}.
In fact, as shown in the inset of Fig. \ref{Fig4} which displays $\alpha_c$ as a function of $g$ (cf Eq. \eqref{croisement}), when $g \rightarrow g_{\perp}^c$, $\alpha_c$ tends to zero.
For $\nu=3$, the LL involved in the crossing studied here are $n_2=2$ and $n_1=1$ so that $Z^c=2$ giving $g_{\perp}^c \approx 49.38$ for $m_{\ast}\approx 0.081 m_e$.
Clearly $\Delta \alpha$ exhibits a (non diverging) peak as $g \rightarrow 49.38$. 
For $\nu=4$, $n_2=3$ and $n_1=1$, there are two non vanishing values of $g$ for which $\alpha_c=0$ : $g_{\perp}^c \approx 74.07$ and $g_{\perp}^c \approx -24.69$ corresponding respectively to $Z^c=3$ and $Z^c=-1$ and also leading to two saturating peaks for 
$\Delta \alpha$.
In fact, one can also see the inset of Fig. \ref{Fig4} as an instability region in the parameter plane $(\alpha,g)$.
The thickness of the lines (whose variations are not visible here) can be measured by the quantity $\Delta \alpha$ at fixed $g$.
However, at fixed $\alpha \equiv \alpha_c = 0$, the width of the instability region should be measured by $\Delta g= \vert g_{\perp} - g_{\perp}^c \vert $, an estimate of which is easily obtained at fixed quantum well width $W$ by inverting Eq. \eqref{delta_Z_LQW}
and by replacing $W_c$ by $W$.

\begin{figure}[t]
\centering
\includegraphics[scale=0.55]{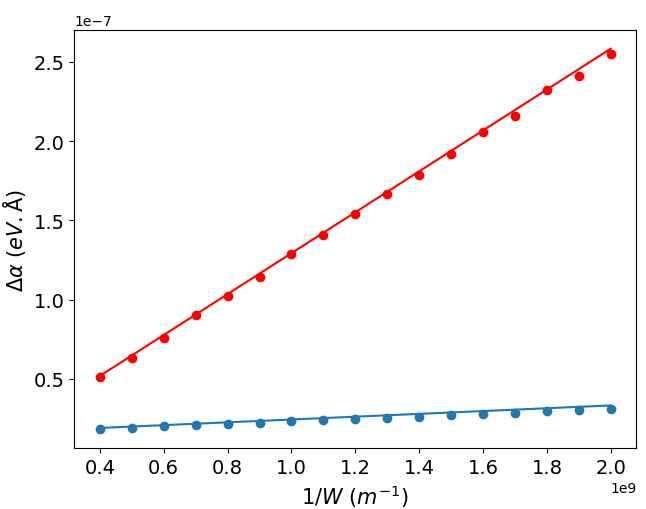}
\caption{$\Delta \alpha$ versus $W^{-1 }$, i.e the inverse of the quantum well width, for an isotropic Zeeman interaction ($g_{\perp}=g_{\parallel}=g = 1.26$) in the case of $\nu=3$ and $\nu=4$ (as before, we fix $m_{\ast}=0.081$ $m_e$). The solid curves correspond to 
the approximate analytical expressions derived in Appendix B (see text), which is linear in  $W^{-1 }$.}
 \label{Fig5}
\end{figure}

Finally, we have looked carefully at the dependence of 
$\Delta\alpha$ 
as a function of                        
the inverse quantum well width $W^{-1}$ for the same conditions as previously, {\em i.e.}, $\nu=3$ or $4$, and $z_0=L_z/2$. As 
observed in Fig. \ref{Fig5}, $\Delta \alpha$ increases linearly with $W^{-1}$ for the chosen parameter range, thus pointing out 
the interest for having the smallest width $W$ as possible to stimulate the instability.  As deduced
from Fig. \ref{Fig4}, the physics here is mostly dictated by the $Q_2$ term. The  found linear dependences 
of $\Delta \alpha$ on $W^{-1}$ is then clear already at the level of Eqs. \eqref{z_0=0} and \eqref{Lz/2}.  In Appendix \ref{AppB}, we 
provide an analytical estimate for the slopes providing the solid curves in Fig. \ref{Fig5}. The excellent agreement found between these estimates
and the numerical results confirms that the $B_x$-Zeeman coupling mechanism is dominant in the present case.

\section{Concluding remarks 
}

We have established that a 2DEG under perpendicular magnetic field can undergo a superradiant instability thanks to an in-plane Zeeman coupling with the photonic field of a cavity resonator. The associated extreme criterion on the quantum well width can be partly relaxed by taking advantage of the singularity of spin-flip transitions, either at Landau level coincidence occurring for specific values of the perpendicular Land\'e factor without Rashba spin-orbit coupling ($\alpha$=0), either at Landau levels crossings produced by finite $\alpha$. As a result, an instability may develop for quantum well widths in the nanoscale for specific values of the effective Land\'{e} factor or of the filling factors. 

Because the Rashba spin-orbit coupling alone gives also rise to a superradiant instability via the coupling to a spatially modulated perpendicular field, two kinds of instability mechanism may in fact work together in the presence of both Rashba and in-plane Zeeman couplings. Nevertheless, the resulting paramagnetic instability still typically occurs close to Landau level crossings/coincidences and requires relatively fine tuning of the model parameters.  Moreover, it turns out that it can be obtained without the cavity, {\em i.e.}, the coupling to the free vacuum field appears sufficient.

Finally, let us discuss the possible experimental realization of our theoretical proposal with the current and available technologies.
First, we should emphasize that state-of-the arts experiments on Landau Polaritons were often focused on the achievement of very large light-matter coupling ratio $\Omega_0/\omega_c$ \cite{Scalari_2012,Maissen_2014}, some times even larger than unity \cite{Bayer_2017}.
In these experiments, the 2DEG was most frequently confined in one or several GaAs  (or AlGaAs/GaAs) quantum wells, with parabolic dispersion and without (or with very small) Land\'e factor and Rashba spin orbit coupling; then, the polaritonic branches measured through spectroscopy as a function of the DC perpendicular magnetic field B (in the range 0 to 10 T typically) were well fitted with the modes calculated from the (non superradiant) Hopfield model, and the large light-matter coupling was achieved in the large filling factor limit (corresponding to small B), in agreement with the original theoretical proposal \cite{Hagenmuller_2010}. 

On the other hand, to realize our proposal, we should work in the integer quantum Hall regime (corresponding to small filling factor), and the quantum wells should be selected in such a way that the 2DEG has both Rashba spin-orbit and Zeeman coupling, so that one can naturally think of (among others) InSb \cite{Morgenstern_2010}, InAs \cite{Maissen_2014}, InP \cite{Hermann1977}, Ge \cite{winkler_book}, or HgCdTe  \cite{Gudina_2022} quantum wells. The values of the parameters that we consider in this work (the Rashba spin-orbit coupling strength, the Land\'e factor, the typical width of the quantum wells W and the values of the filling factor) are typical for realistic structures.  
Moreover, a deviation from the Hopfield model towards the Dicke model has been measured in strained Ge and InSb quantum wells  \cite{keller2020landau} where Rashba and Zeeman couplings are known to be important. According to our theory,
to push the system towards the superradiant phase, the physical parameters should be fine-tuned towards Landau Level crossings. 

Thus, the {\it in-situ} tunability of the Rashba spin-orbit strength $\alpha$ thanks to an applied gate voltage would be an interesting option \cite{Nitta_1997,Chirolli_2012, Rossi_2022}.
The possibility of varying the Land\'e g-factor, although less obvious, could also be realized in 2DEGs where it is B (or energy) dependent, as a consequence of the non parabolicity of the band and/or the exchange energy \cite{scriba_1993,Failla_2016}, but accurate predictions
require complementary calculations.
Moreover, the size of the instability regions in the parameter space $(\alpha,g)$ that we predict here and which eventually determines how fine-tuned should be $\alpha$ and/or $g$ for a given quantum well width is such that $\Delta \alpha / \alpha \sim \Delta g/ g \sim 10^{-4}$ to $10^{-5}$ (for $W=1$ nm as appearing in Eq. (\ref{delta_Z_LQW}) and in Fig. \ref{Fig4}). 
It is worth noting that such a ratio is much larger than analogous ones appearing in other theoretical proposals based on van Hove singularities\cite{Guerci2020, Guerci2021, Sanchez2021}, but still small enough to represent an experimental issue, since disorder and impurities broaden the Landau Levels. 

Out of scope of the present paper, other ingredients which deserve future investigations are the possibly detrimental influence of the disorder, as well as the possibly beneficial influence of the Coulomb interaction which is likely to further soften the excitations due to the excitonic effect. 
Finally, in this work, we considered a simple cavity geometry with a simplified description of its field, focusing on the transverse electric modes propagating in the $x$ direction.
A more realistic description of the resonators used in the state-of-the arts experiments, like the single \cite{rajabali_2022} or arrays \cite{keller2020landau} of complementary split ring resonator(s), may change our results quantitatively.

%

\begin{acknowledgments}
This work has been supported by the French National Research Agency in the framework of the "Investissements d'avenir" program (ANR-15-IDEX-02).
\end{acknowledgments}
 
\appendix

\section{Susceptibility and polariton modes}

\label{AppA}


In this Appendix, we provide the technical details leading to the polariton mode Eqs. \eqref{Lz/2}-\eqref{z_0=0}.
We first need to evaluate the response function 
$\mathcal{Q}_{kl}({\bf r}, {\bf r'},\omega)$ of the 2DEG in the presence of Zeeman and Rashba spin-orbit couplings.  The electronic current density is obtained from Hamiltonian \eqref{2DEG} as ${\bf j}= -c \, \partial H/\partial {\bf A}$ and thus reads
\begin{eqnarray}
{\bf j} =  -\frac{e}{m_{\ast}} {\bf p} + e \alpha {\bf u}_z \times \boldsymbol{\sigma}  - \frac{e^2  }{m_{\ast} } \frac{{\bf A}}{c} - \frac{ \mu_B c}{2} \boldsymbol{\nabla}\times  \left(\hat{g} \boldsymbol{\sigma} \right). \hspace*{0.5cm}
\end{eqnarray}
Taking ${\bf A} = {\bf A}^{\mathrm{ext}} + \delta {\bf A}$, the term proportional to $\delta {\bf A}$  gives the diamagnetic contribution to the response, while the rest should be plugged in the Kubo formula to get
\begin{align}
&\mathcal{Q}_{kl}({\bf r},{\bf r'}, \omega) = \frac{n_{e} e^2}{m_{\ast}} \delta({\bf r}-{\bf r'} ) \, \delta_{kl} \nonumber \\
& -   \frac{i}{\hbar} \int dt \, e^{i(\omega + i 0^{+}) t} \left\langle \left[   j_{k} ( {\bf r},t) , j_{l} ( {\bf r'},0)\right]   \right\rangle_{0},
\label{Kubo}
\end{align}
where the subscript $0$ means the quantum average in the state before the field $\delta {\bf A}$ is turned on. 
 The 2DEG eigenstates labeled by $\ell \equiv (n,\lambda)$ and the momentum $k_x$ are given in the Landau gauge  ${\bf A}_{\mathrm{ext}} = ( - B y, 0, 0)$ by
\begin{align}
\label{xi}
\langle {\bf r} | \eta \rangle = \frac{e^{ik_x x} }{\sqrt{L_x}} \zeta(z) \begin{pmatrix}
 \phi_{n-1}(y-k_x l_B^2) \sin\theta_{\ell} \\  
i \, \phi_{n}(y-k_x l_B^2)\cos\theta_{\ell}
\\
\end{pmatrix}
\end{align}
with the angles $\theta_{\ell}$ defined in the main text and
\begin{align}
\phi_n(y) = \frac{e^{-y^2/(2 l_B^2)}}{\sqrt{2^n n! l_B \sqrt{\pi}}} H_n
\left(\frac{y}{l_B} \right),\label{eq:Hermite}
\end{align}
where $H_n$ is the Hermite polynomial of degree $n$ [here $\eta=(\ell, k_x)$]. $\zeta(z)$ is the wave function of the lowest subband corresponding to the transverse confinement of the electrons in the quantum well.  Crucially, we consider that the quantum well width is the smallest length scale in the problem, so we associate $\zeta^2(z)=\delta(z-z_0)$ in all cases when it must be integrated with a smooth function. On the contrary, when we enconter the integral $\int\delta^2(z-z_0)\,dz=\int\zeta^4(z)\,dz$, we associate it with the inverse quantum well width~$1/W$.

We are only interested  in the response function in the $y$ direction, {\em i.e.}, $k=l$ in Eq. \eqref{Kubo}. Furthermore, considering the  Fourier transform \eqref{TF}, we obtain
\begin{eqnarray}
Q( q_x,z,z', \omega) &=& \frac{n_{e} e^2}{m_{\ast}} \,\delta(z-z_0) \,\delta(z'-z_0)\nonumber \\  
&&{}+\frac{2}{L_x L_y} 
\sum_{\ell \leq \nu < \ell'} 
 \sum_{k_x,k_x'}  (\epsilon_{\ell'} - \epsilon_{\ell})\nonumber\\
&&\quad{}\times \frac{  
\langle \eta |j_y(q_x,z) | \eta' \rangle 
\langle \eta' |j_y(q_x,z') | \eta \rangle 
}{\left(\hbar \omega\right)^2 -(\epsilon_{\ell'} - \epsilon_{\ell})^2} , 
\nonumber\\ \label{Qend}
\end{eqnarray}
where the current matrix elements  are evaluated as
\begin{eqnarray}
\bra{\eta} j_{y} (q_x,z) \ket{\eta'}  
&=& i  \delta_{k_x-q_x,k_x'} \nonumber\\
&&{}\times\left[ \mathcal{A}_{q_x}^{\ell \ell'} \delta(z-z_0) 
+\mathcal{B}_{q_x}^{\ell \ell'}  \delta'(z-z_0)
\right],\nonumber\\
\end{eqnarray}
with $\mathcal{A}_{q_x}^{\ell' \ell} $ and $\mathcal{B}_{q_x}^{\ell' \ell}$ given in Eqs. \eqref{defA}--\eqref{defB}. Performing the sums over $k_x,k_x'$ in Eq. \eqref{Qend}, we finally get Eq. \eqref{susceptibilite}. 

Considering a vector potential  ${\bf A}={\bf u}_y A_{y}(z)e^{i q_x x}$, Eq.~\eqref{eq} then becomes
\begin{align}
\label{Discontinuite}
&  \kappa^2  A_y(z) - \partial^{2}_z  A_y (z)  \nonumber \\
& = \frac{4  \pi}{c^2} \delta(z-z_0) Q_{0}(q_x,\omega) \int \mathrm{d}z'~ \delta(z'-z_0) A_y(z')  \nonumber \\
& + \frac{4  \pi}{c^2} \delta(z-z_0)Q_{1}(q_x,\omega)\int \mathrm{d}z'~ \delta'(z'-z_0) A_y(z') \nonumber \\
& + \frac{4  \pi}{c^2} \delta'(z-z_0)Q_{1}(q_x,\omega) \int \mathrm{d}z'~ \delta(z'-z_0) A_y(z') \nonumber \\
& + \frac{4  \pi}{c^2} \delta' (z-z_0) Q_{2}(q_x,\omega)\int \mathrm{d}z'~ \delta'(z'-z_0) A_y(z').
\end{align}
For $\partial^{2}_z  A_y (z)$ to be as singular as $\delta'(z-z_0)$, $A_y(z)$ itself must have a jump at $z=z_0$.
Hence, we search a solution under the form 
\begin{align}
 A_{y} (z) =  \begin{cases}
A_1   \sinh \kappa z, \quad z<z_0,\\
A_2 \sinh(\kappa L_z- \kappa z), \quad z> z_0, \label{form}
\end{cases}
\end{align}
which satisfies the equation away from $z=z_0$ and obeys the boundary conditions $A_y(0)=A_y(L_z)=0$. Substituting \eqref{form} into Eq. \eqref{Discontinuite}, we encounter two singular integrals involving the Heaviside step function~$\theta(z)$ and the Dirac $\delta(z)=d\theta(z)/dz$:
\begin{align}
&\int\delta(z)\,\theta(z)\,dz=\frac{1}2\int\frac{d\theta^2(z)}{dz}\,dz=\frac12,\\
&\int\delta'(z)\,\theta(z)=-\int\delta^2(z)\,dz=-\frac{1}W.
\end{align}
The latter expression is our definition of the quantum well width~$W$ -- the typical scale over which the electron wave function is spread [see the discussion after Eq.~(\ref{eq:Hermite})]. The jump in the vector potential at $z=z_0$, expressed by Eq.~(\ref{form}), corresponds to a contribution to $B_x\propto\delta(z-z_0)$, which couples to the in-plane magnetization $M_x\propto\delta(z-z_0)$; the resulting interaction energy is proportional to $\int\delta^2(z-z_0)\,dz=1/W$.

Equating the coefficients in Eq.~\eqref{Discontinuite} in front of the terms  $ \delta(z-z_0) $ and $ \delta'(z-z_0) $, we obtain a linear system for the amplitudes $A_1$ and $A_2$ 
\begin{align}
\begin{pmatrix}
\mathcal{M}_{11} &   \mathcal{M}_{12} \\
\mathcal{M}_{21} &   \mathcal{M}_{22}
\end{pmatrix}
\begin{pmatrix}
A_1 \\
A_2
\end{pmatrix}
=0 , \label{mat}
\end{align}
where
\begin{eqnarray}
\mathcal{M}_{11} &=&   \kappa  \left(1+\frac{2 \pi }{c^2}Q_1 \right)\cosh \kappa  z_0
\nonumber \\
&& {} - \frac{4  \pi}{c^2}  \left( \frac{Q_0}{2}+ \frac{Q_1}{W}\right)\sinh \kappa  z_0,
  \nonumber \\
\mathcal{M}_{12} &=& \kappa  \left(1-\frac{2 \pi }{c^2}Q_1 \right)\cosh(\kappa L_z-\kappa z_0)
   \nonumber \\
&& 
{}- \frac{4  \pi}{c^2}  \left( \frac{Q_0}{2}- \frac{Q_1}{W}\right)
\sinh( \kappa  L_z- \kappa z_0),
 \nonumber \\
\mathcal{M}_{21} & =& 
   \left( 1- \frac{4 \pi}{c^2} \frac{Q_2}{W}
- \frac{2 \pi}{c^2} Q_1
\right)\sinh \kappa  z_0
\nonumber \\
&& {} + \frac{2  \pi}{c^2} \kappa Q_2 \cosh \kappa  z_0  ,  \nonumber \\
\mathcal{M}_{22} &=& - \left( 1- \frac{4 \pi}{c^2} \frac{Q_2}{W}
+ \frac{2 \pi}{c^2} Q_1
\right) \sinh(\kappa L_z-\kappa z_0) 
  \nonumber \\
&& {}- \frac{2  \pi}{c^2} \kappa  q_2 \cosh( \kappa  L_z-\kappa z_0) . \nonumber
\end{eqnarray}
To ensure the existence of solutions to the linear system \eqref{mat}, the matrix determinant must vanish. This leads to Eq. \eqref{Lz/2} for $z_0= L_z/2$, and to Eq. \eqref{z_0=0} for $z_0 =0$ after taking into account that $\kappa W \ll 1$.

\section{Analytical estimate for the quantum well width and for the ``Superradiant'' phase width}
\label{AppB}

In this Appendix, we provide some analytical simplifications, which are helpful to determine Eq. \eqref{delta_Z_LQW} of the main text and for the analysis of the instability regions. Firstly, close to either a level crossing or a level coincidence, we can realize from Eqs. \eqref{Q0}--\eqref{Q2} that at $\omega=0$ the $Q_i$ are mostly given by a single diverging term in the sums over $(\ell,\ell')$. 
For instance, when $\epsilon_{n_1,+} \approx  \epsilon_{n_2,-}$, we can approximate the $Q_2$ contribution as (we do not specify the superscript of $\mathcal{B}_{q_x}$ for convenience)
\begin{eqnarray}
Q_2(q_x,0) \approx \frac{1}{\pi l_B^2} \frac{\mathcal{B}_{q_x}^2}{\left|\epsilon_{n_1,+}-\epsilon_{n_2,-} \right|}, \label{Q2ap}
\end{eqnarray}
and similarly for the other contributions.
As a result, we have  $ \left[Q_{1}(q_x,0)\right]^2 -  Q_{0}(q_x,0)  Q_{2}(q_x,0) \approx 0$, which means that the latter combination plays a negligible role in Eqs. \eqref{Lz/2}--\eqref{z_0=0} for $\omega=0$.

Simplifications of the dipole matrix elements also take place when considering  $\nu=n_1+n_2 \gg 1$ (the order-of-magnitude estimate is expected to be valid for $\nu \sim 1$ as well). Indeed, we then use the asymptotic expression for the generalized Laguerre polynomials with large index in terms of the Bessel function of the first kind $J$. Eq. \eqref{Theta} becomes
\begin{align} \label{bessel}
\Theta_{n_2}^{n_1} & \simeq  S_{n_2}^{n_1} \, J_{|n_1 - n_2|} \left( \sqrt{\nu} l_B q_x \right).
\end{align}
Moreover, to the leading order in $1/\sqrt{\nu}$, we get the relation between the angles, $\theta_{n_2,-}=\theta_{n_1,+}+\pi/2+ \mathcal{O}(1/\nu)$, which expresses the approximate orthogonality of the spin part of the wave functions for $|n_1-n_2| \ll n_1, \, n_2$ and $\lambda_1=-\lambda_2$. 
Thus,
\begin{eqnarray}
\mathcal{B}_{q_x}^2 \approx
\left(\frac{g_{\parallel}  \mu_B c }{2} \right)^2 \left(
  \Theta_{n_1-1}^{n_2}\sin^2 \theta
+  
 \Theta_{n_1}^{n_2-1}\cos^2 \theta
\right)^2, \nonumber
\end{eqnarray}
 with $\theta \equiv \theta_{n_1,+}$. \\
Let's first focus on the case where $\alpha=0$ and $g_{\perp} \rightarrow g_{\perp}^c$ to explain Eq. (\ref{delta_Z_LQW}).
Firstly, one has:
\begin{align} \label{delta_energie_gc}
\left|\epsilon_{n_1,+}-\epsilon_{n_2,-} \right| \approx \hbar \omega_c \left ( \frac{m_{\ast}}{2 m_e}\right ) \vert g_{\perp}- g_{\perp}^c\vert.
\end{align}
Secondly, since $\theta=\theta_{n_1,+}\rightarrow \pi/2$, one has $\mathcal{B}_{q_x}^2 \approx
\left(\frac{g_{\parallel}  \mu_B c }{2} \right)^2 
  (\Theta_{n_1-1}^{n_2})^2$. Moreover, since $\theta_{n_2,-}-\theta_{n_1,+} \approx \pi/2$, the terms $\mathcal{A}_{q_x}^{\ell' \ell} $ in Eq. \eqref{defA} tends to 0,
so both the $Q_{0}$ and $Q_{1}$ terms are very small so that both Eq. \eqref{Lz/2} and  \eqref{z_0=0} reduce to 
$\frac{Q_2(q_x,0)}{W} = \frac{c^2}{4 \pi}$. Using expression \eqref{Q2ap}, it appears that such an  equation starts to have solutions in the parameter space as early as  the  maximum of $\mathcal{B}_{q_x}^2$ (which is the maximum of $(\Theta_{n_1-1}^{n_2})^2$) is reached when varying $q_x$.  As soon as $n_2-n_1 \neq -1$ ($n_2-n_1=-1$ would correspond to the case $g_{\parallel}=g_{\perp}^c=0$ for isotropic Zeeman coupling), such a maximum appears for finite $q_x$ since the Bessel function appearing in Eq. \eqref{bessel} will have a non zero index.
Modulo some unimportant numerical factor given by the value of such a maximum, we finally obtain Eq. \eqref{delta_Z_LQW} of the main text using both Eqs. \eqref{Q2ap} and \eqref{delta_energie_gc}. \\
 From now on, we consider a situation where the Rashba spin-orbit coupling $\alpha$ is close to  $\alpha_c$, its value at the crossing, that we consider not too close to 0, for fixed $g_{\perp}$ (away from $g_{\perp}^c$).

From Eq. \eqref{croisement}, we get that for $\nu \gg 1$, $\alpha_c$  is such that
\begin{align}
m_{\ast}\alpha_cl_B \approx \sqrt{\frac{(n_1-n_2)^2 - (1- Z)^2}{4 \nu }} \ll 1 .
\end{align}

When $\alpha$ is detuned away from $\alpha_c$ (keeping both the magnetic field and the filling constant), we have
\begin{eqnarray}
\label{omegaLL}
\left|\epsilon_{n_1,+}-\epsilon_{n_2,-} \right| \approx \frac{2 \hbar}{l_B}  \sqrt{\nu} \sqrt{1- \left|\frac{ 1- Z}{n_1-n_2} \right|^2} \, \vert \alpha - \alpha_c \vert . \hspace*{0.8cm}
\end{eqnarray}
  Using Eqs. \eqref{Q2ap} and \eqref{omegaLL} and considering that $Q_2$ provides the main contribution, Eq. \eqref{Lz/2} for $z_0=L_z/2$ at $\omega=0$ then yields 
\begin{align}
\label{deltaalpha}
\vert \alpha - \alpha_c \vert   \simeq \frac{1}{\hbar}
\frac{4  \pi}{c^2} \frac{n_e}{\nu^{3/2}}
\frac{l_B}{W }    
\frac{|n_1-n_2| \mathcal{B}_{q_x}^2 }{ \sqrt{(n_1-n_2)^2 - (1-Z)^2 }} .
\end{align}
Note that this equation is valid for $g_{\perp}$ away from $g_{\perp}^c$ so that the denominator of the LHS does not vanish, which corresponds to $\vert \alpha_c \vert > 0$.
Moreover and similarly to what was done before, this equation starts to have solutions when  the  maximum of $\mathcal{B}_{q_x}^2$ is reached. The typical value for $q_x$ associated to this maximum depends on $\nu$. By Plugging it in Eq. \eqref{deltaalpha}, we then get an approximate estimation for the superradiant phase width $\Delta \alpha=|\alpha -\alpha_c|$.
For $\nu =3$, the  maximum of $\mathcal{B}_{q_x} ^2$ is at $q_x=0$ for a large range of positive $g$ (except when $\cos \theta \to 0$). In contrast, for $\nu=4$, it is typically reached for $ l_B q_x \approx 1$. The corresponding derived $\Delta \alpha$ are represented by the solid lines in Fig. \ref{Fig5}.




\bibliography{paper_biblio}

\end{document}